\documentclass{article}
\usepackage{multicol}

\begin{multicols}{2}
\title{Nonlocal strain gradient exact solutions for functionally graded
inflected nano-beams}
\author{
       A. Apuzzo\\
       Department of Civil and Mechanical Engineering\\
       University of Cassino and Southern Lazio\\
       a.apuzzo@unicas.it
     \and
       R. Barretta\\
       Department of Structures for Engineering and Architecture\\
       University of Naples Federico II\\
       rabarret@unina.it\\
     \and
     	S. A. Faghidian\\
	Department of Mechanical Engineering\\
	Islamic Azad University\\
	faghidian@gmail.com\\
      \and
      R. Luciano\\
      Department of Civil and Mechanical Engineering\\
      University of Cassino and Southern Lazio\\
      luciano@unicas.it\\
     \and
      F. Marotti de Sciarra\\
      Department of Structures for Engineering and Architecture\\
       University of Naples Federico II\\
       marotti@unina.it\\      
}
\date{}
\end{multicols}

\usepackage{setspace}   %Allows double spacing with the \doublespacing command

\usepackage{amssymb}
\usepackage{amscd}
\usepackage{amsthm}
\usepackage{mathtools}% includes amsmath
\usepackage{amsfonts}
\usepackage{longtable}
\usepackage[all,knot,poly]{xy}
\usepackage{latexsym}
\usepackage{graphicx}
\usepackage{epsfig}
\usepackage{epstopdf}
\usepackage[dvipsnames,usenames]{color}
\usepackage[colorlinks=true,
pdfstartview=FitV, linkcolor=cyan,
citecolor=cyan, urlcolor=cyan]{hyperref}
\usepackage{enumerate}
\usepackage{enumitem}
\usepackage{subcaption}
\usepackage{balance}
\input{tcilatex}

\begin{document}

%%\begin{frontmatter}

\maketitle

%%\doublespacing

\begin{abstract}
The size-dependent bending behavior of nano-beams is investigated
by the modified nonlocal strain gradient elasticity theory.
According to this model, the bending moment is expressed 
by integral convolutions of elastic flexural curvature 
and of its derivative with a bi-exponential averaging kernel.
It has been recently proven that such a relation is equivalent 
to a differential equation, involving bending moment and flexural curvature fields, 
equipped with natural higher-order boundary conditions of constitutive type.
The associated elastostatic problem of a Bernoulli-Euler functionally graded nanobeam 
is formulated and solved for simple statical schemes of technical interest.
An effective analytical approach is presented and exploited to
establish exact expressions of nonlocal strain gradient transverse displacements
of doubly clamped, cantilever, clamped-pinned and pinned-pinned
nano-beams, detecting thus also new benchmarks for numerical analyses.
Comparisons with results of literature, corresponding to 
selected higher-order boundary conditions are provided and discussed.
The considered nonlocal strain gradient model
can be advantageously adopted to characterize  
scale phenomena in nano-engineering problems.
\end{abstract}
%%%%%%%%%%%%%%%%%%%%%%%%%%%%%%%%% 


%%\end{frontmatter}


\doublespacing

\section{Introduction}
\noindent
In recent years new and multifunctional materials have been introduced
requiring the consideration of small length scales \cite{Patti,Acierno}. 
From an engineering standpoint, the realization of nano-actuators, nano-sensors, 3D
printings and structural components for micro- and nano-systems has become an important topic, see e.g. the
review papers \cite{Kumar,Wang} and the contributions \cite{Sedighi3}-%
\cite{Mojahedi}. Since mechanical properties are size-dependent at the
nanoscale, the study of size effects on the behaviour of nano-beams is an
important area of research. It is well-known that the classical continuum
theory neglects structural phenomena that are important at small-scales
\cite{GreIJSS2010}-\cite{Karami}.
Accordingly new non-classical continuum theories can be adopted to model the
size-dependency of micro- and nano-scale structures such as strain-driven
and stress-driven nonlocal elasticity 
\cite{FMdSIJSS2008}-\cite{JCOMPB2017}.
\noindent
The strain-driven nonlocal elastic model has been introduced by Eringen \cite%
{Eringen83} and the nonlocal stress is obtained\ in terms of the integral
convolution of the elastic strain with a smoothing kernel. Such a model
cannot be adopted to assess size-effects in nanomechanics, as proved in \cite%
{Romano17} and agreeded in several papers such as \cite{Barati,Sahmani}. 
Actually, the ill-posedness of Eringen's strain-driven purely nonlocal
elastic model for bounded domains relies on the fact that the constitutive
boundary conditions associated with Eringen's integral law conflict with
equilibrium requirements, see \cite{Romano16,Barretta18a}.
\noindent
The stress-driven nonlocal elastic model for nano-beams has been provided in 
\cite{Romano17a} and the nonlocal elastic strain is obtained\ in terms of
the integral convolution of the stress with a smoothing kernel. This model
leads to well-posed nonlocal problems for structures defined on bounded
domains, see e.g. \cite{Apuzzo}-\cite{Barretta18c}.
\noindent
A recent approach proposed in \cite{Lim} merges Eringen's nonlocal
strain-driven integral law with the strain gradient elasticity to obtain a
higher-order nonlocal theory. The nonlocal stress is defined as the sum of
two integral terms: the former is Eringen's convolution between the
elastic strain and a smoothing kernel depending on a nonlocal parameter and
the latter is the derivative of the convolution of the elastic strain
gradient with a smoothing kernel depending on a nonlocal parameter. If the
bi-exponential function is considered for the smoothing kernel, the solution
of the nonlocal problem is achieved, in literature, by replacing the
integral law with the differential law which is considered to be equivalent
to the integral relation. The main problem is that such a differential
equation is of higher-order than the one of the classical local problem. As
a consequence, additional (so-called) higher-order boundary conditions have
to be imposed to solve the nonlocal strain gradient elastostatic problem. In
literature, two different choices are provided consisting in imposing
higher-order kinematic \cite{Li} or static \cite{Xu} boundary conditions
pertaining to the strain gradient theory. It is worth noting that the
nonlocal structural response is particularly affected by these choices.
\noindent
A definitive answer to the discussion on the higher-order boundary
conditions to be imposed in order to solve the elastostatic problem of
nonlocal strain gradient nano-beams has been provided in \cite{Barretta18d}.
It is proved that such boundary conditions follow from the nonlocal
strain gradient integral law and turns out to be of constitutive kind. It is
also shown in \cite{Barretta18d} that the differential problem, equipped
with the correct constitutive boundary conditions, leads to well-posed
elastostatic problems for bounded nano-structures.
\noindent
The nonlocal strain gradient elastic model is utilized in the present paper
to address the size-dependent static behavior of inflected nano-beams of
technical interest. Small-scale effects are exhibited by nano-beams for any
boundary condition.
\noindent
In particular, closed form solutions for the nonlocal strain gradient model
are given for cantilever, simply-supported, clamped-pinned and
fully-clamped FG nano-beams subject to a uniform load. It is
shown that nano-beams exhibit softening and stiffening structural behaviors
for increasing nonlocal and gradient parameters respectively, so that the
nonlocal strain gradient law provides an effective approach to design a
wide class of nanodevices. If the gradient parameter vanishes, the constitutive boundary conditions of the nonlocal strain gradient
law coincide to the ones of Eringen's nonlocal integral law given
in \cite{Romano17}.
\noindent
Numerical analyses are provided as benchmarks for applications
and experiments involving inflected nano-beams.

\section{Nonlocal strain gradient model for Bernoulli-Euler nano-beams}
\noindent
Nonlocal strain gradient (NSG) model of elasticity for inflected nano-beams,
according to the model proposed in \cite{Lim}, is formulated by expressing
the bending moment $M$ in terms of elastic flexural curvature $\chi _{el}$
and of its derivative $\partial _{x}\chi _{el}$ as

\begin{equation}
\begin{array}{l}
M\left( x,\lambda _{0},\lambda _{1},l\right) =\left( \alpha _{0}\ast \left(
K\cdot \chi _{el}\right) \right) \left( x,\lambda _{0}\right) -l^{2}\partial
_{x}\left( \alpha _{1}\ast \left( K\cdot \partial _{x}\chi _{el}\right)
\right) \left( x,\lambda _{1}\right)  \\ 
\vspace{-4pt}\cr
=\dint_{0}^{L}\alpha _{0}\left( x-y,\lambda _{0}\right) \cdot \left(
K\cdot \chi _{el}\right) \left( y\right) dy-l^{2}\partial _{x}\dint_{0}^{L}\alpha _{1}\left( x-y,\lambda_{1}\right) \cdot \left( K\cdot \partial _{x}\chi _{el}\right) \left(
y\right) dy.%
\end{array}
\label{g1}
\end{equation}%
We consider a straight beams of length $L$, the $x$-coordinate is taken
along the length of the nano-beam with the $y$-coordinate along the
thickness and the $z$-coordinate along the width of the nano-beam. 

The local
elastic bending stiffness is $K=I_{E}$ where $I_{E}=\int_{\Omega }E\left(
y\right) dA$ is the second moment of the field of Young elastic moduli $E$
along the $y$ bending axis.
\noindent
The smoothing kernels $\alpha _{0}$ and $\alpha _{1}$ depend on two
non-dimensional non-local parameters $\lambda _{0}$ \TEXTsymbol{>} 0 and $%
\lambda _{1}$ \TEXTsymbol{>} 0 . The characteristic length $l\geq 0$ has
been introduced to make dimensionally homogeneous the convolutions in Eq. (%
\ref{g1}).
\noindent
Following \cite{Lim} - \cite{Li}, we consider that the nonlocal parameters
are coincident, i.e. $\lambda $ $:=\lambda _{0}$ $=\lambda _{1}$, and the
kernels $\alpha _{0}$ and $\alpha _{1}$ are coincident with the
bi-exponential averaging function given by%
\begin{equation}
\phi (x,c)=\dfrac{1}{2 c}\exp \left( -\dfrac{\left\vert x\right\vert 
}{c}\right) ,  \label{g2}
\end{equation}%
being $c=\lambda L$ the characteristic length of Eringen nonlocal
elasticity. The bi-exponential  function fulfils positivity, symmetry,
normalisation and impulsivity, 
\noindent
Introducing the following fields \cite{Lim}%
\begin{equation}
\begin{array}{l}
M_{0}\left( x,c\right) =\dint_{0}^{L}\phi \left( x-y,c\right) \cdot
\left( K\cdot \chi _{el}\right) \left( y\right) dy \\ 
\vspace{-4pt}\cr
M_{1}\left( x,c,l\right) =l^{2}\partial _{x}\dint_{0}^{L}\phi \left(
x-y,c\right) \cdot \left( K\cdot \partial _{x}\chi _{el}\right) \left(
y\right) dy,%
\end{array}
\label{g3}
\end{equation}%
the nonlocal strain gradient elastic law Eq. (\ref{g1}) can then be
rewritten as%
\begin{equation}
M\left( x,c,l\right) =M_{0}\left( x,c\right) -\partial
_{x}M_{1}\left( x,c,l\right) .  \label{g4}
\end{equation}
\noindent
Since the bi-exponential averaging kernel Eq. (\ref{g2}) is the Green
function on the whole real axis of Helmholtz linear differential operator $%
\mathcal{L}_{c}=1-c^{2}\partial _{x}^{2}$, the differential constitutive
relation ensuing the nonlocal strain gradient law Eq. (\ref{g4}) can
be written as%
\begin{equation}
\left( K\cdot \chi _{el}\right) \left( x\right) -l^{2}\partial
_{x}^{2}\left( K\cdot \chi _{el}\right) \left( x\right) =M\left(
x,c,l\right) -c^{2}\ \partial _{x}^{2}M\left( x,c,l\right) .
\label{g5}
\end{equation}
\noindent
It is proved in \cite{Barretta18d} that the nonlocal strain gradient
integral relation Eq. (\ref{g4}) is not equivalent to the differential law
Eq. (\ref{g5}) for nano-beams defined on a bounded interval $[0,L]$. In fact
suitable boundary conditions must be prescribed to ensure the constitutive
equivalence according to the following proposition proved for the first time
in \cite{Barretta18d}.
\noindent
\textbf{Proposition 1 - Constitutive equivalence property. }The nonlocal
strain gradient constitutive law Eq. (\ref{g4}) equipped with the
bi-exponential kernel Eq. (\ref{g2})%
\begin{equation}
M\left( x,c,l\right) =M_{0}\left( x,c\right) -\partial
_{x}M_{1}\left( x,c,l\right) ,  \label{g6}
\end{equation}%
with $x\in \left[ 0,L\right] $, is equivalent to the differential relation
Eq. (\ref{g5})%
\begin{equation}
\left( K\cdot \chi _{el}\right) \left( x\right) -l^{2}\partial
_{x}^{2}\left( K\cdot \chi _{el}\right) \left( x\right) =M\left(
x,c,l\right) -c^{2}\ \partial _{x}^{2}M\left( x,c,l\right) 
\label{g7}
\end{equation}%
subject to the following two constitutive boundary conditions (CBC) %
\begin{equation}
\left\{ 
\begin{array}{l}
\begin{array}{l}
\partial _{x}M\left( 0,c,l\right) =\dfrac{1}{c}M\left(
0,c,l\right) +\dfrac{l^{2}}{c^{2}}\partial _{x}\left( K\cdot \chi
_{el}\right) \left( a\right)  \\ 
\vspace{-4pt}\cr
\partial _{x}M\left( L,c,l\right) =-\dfrac{1}{c}M\left(
L,c,l\right) +\dfrac{l^{2}}{c^{2}}\partial _{x}\left( K\cdot \chi
_{el}\right) \left( L\right) .%
\end{array}%
\end{array}%
\right.   \label{g8}
\end{equation}

$\hfill \blacksquare $

\section{\noindent \textbf{Static flexural analysis}}
\noindent
To demonstrate the differences in the flexural results of the proposed
nonlocal model and the existing models in the literature, a flexural beam
problem with the Bernoulli-Euler kinematics is investigated. The
well-established differential and classical boundary conditions of static
equilibrium may be expressed as 
\begin{equation}
\begin{array}{l}
{\partial _{x}^{2}M\left( x\right) =q} \\ 
{\left. M \left(\partial _{x} \delta v\right) \right\vert
_{0,L}=\left. \left( \partial _{x}M\right) \delta v\right\vert
_{0,L}=0.}%
\end{array}
\label{GrindEQ__1_}
\end{equation}%
Assuming coincidence of elastic and total flexural curvatures
$\,\chi _{el}=\chi\,$ and 
employing the differential equation of equilibrium in addition to the
condition of kinematic compatibility $\,\chi=\partial _{x}^{2}v\,$, the
bending moment introduced in Eq. (\ref%
{g7}) can be expressed as 
\begin{equation}
M\left( x\right) =c^{2}q+\left( K\cdot \partial
_{x}^{2}v\right) \left( x\right) -l^{2}\partial _{x}^{2}\left( K\cdot
\partial _{x}^{2}v\right) \left( x\right)  \label{GrindEQ__2_}
\end{equation}%
In terms of deflection of nano-beam,
the differential governing equation takes the form 
\begin{equation}
\partial _{x}^{2}\left( K\cdot \partial _{x}^{2}v\right) \left( x\right)
-l^{2}\partial _{x}^{4}\left( K\cdot \partial _{x}^{2}v\right) \left(
x\right) =q-c^{2}\cdot \partial _{x}^{2}q  \label{GrindEQ__3_}
\end{equation}%
equipped with the classical boundary conditions Eq. \eqref{GrindEQ__1_}$%
{}_{2}$ and the constitutive boundary conditions Eq. (\ref{g8}).
\noindent
Employing the proposed nonlocal model, the exact solutions of the
flexural analysis are derived for nano-beams with customary boundary
conditions and subject to distributed loads. For a uniformly distributed
load $q_{0}$, analytical solution of the static equations governing the
flexural deflection of the nano-beam, Eq. \eqref{GrindEQ__3_}, can be
expressed by 
\begin{equation}
v\left( x\right) =\frac{q_{0}x^{4}}{24K}+\Gamma _{1}\mathrm{exp}\left( \frac{%
x}{l}\right) l^{4}+\Gamma _{2}\mathrm{exp}\left( -\frac{x}{l}\right)
l^{4}+\Gamma _{3}x^{3}+\Gamma _{4}x^{2}+\Gamma _{5}x+\Gamma _{6}
\label{GrindEQ__4_}
\end{equation}%
where $\Gamma _{k}\left( k=1,\ldots ,6\right) $ are unknown constants to be
determined by suitable boundary conditions including the well-known
classical boundary conditions Eq. \eqref{GrindEQ__1_}${}_{2}$ and the
constitutive boundary conditions Eq. (\ref{g8}).
\noindent
The exact analytical solutions for flexure problem of Bernoulli-Euler
nano-beams will be also compared with the counterpart results introduced in
the literature by prescribing different kind of higher-order constitutive
boundary conditions. 
Assuming coincidence of elastic and total flexural curvatures
$\,\chi _{el}=\chi\,$, the higher-order constitutive boundary conditions
adopted in literature may be expressed as either of the following conditions
\cite{Li}.
\noindent
The first constitutive boundary conditions, referred to as
Kinematic Higher-Order Boundary Conditions (KHOBC), 
are expressed by the vanishing of 
the flexural curvature at the end cross-sections 
\begin{equation}
\left. \chi\right\vert _{x=0,L}=\left. \partial _{x}^{2}v\right\vert
_{x=0,L}=0  \label{GrindEQ__5_}
\end{equation}
\noindent
The second constitutive boundary conditions, referred to as 
Static Higher-Order Boundary Conditions
(SHOBC), are expressed by the vanishing of 
the derivative of flexural curvature at the end
cross-sections 
\begin{equation}
\left. \partial _{x}\chi\right\vert _{x=0,L}=\left. \partial
_{x}^{3}v\right\vert _{x=0,L}=0  \label{GrindEQ__6_}
\end{equation}
\noindent
Bernoulli-Euler nano-beam with four different sets of boundary
conditions including doubly clamped, clamped-simply supported, simply
supported and cantilever nano-beams under a uniform load $q_{0}$ are
examined. 

While the set of classical boundary conditions are enforced to the
beam ends according to local Bernoulli-Euler beam theory, the constitutive
boundary conditions are imposed to corresponding beam ends in the proposed
nonlocal model as Eq. (\ref{g8}) yielding in the deflection \textit{v }and
the higher-order constitutive boundary conditions as Eq. \eqref{GrindEQ__5_}
and \eqref{GrindEQ__6_} are enforced for the counterpart models resulting in
the deflections $v^{\mathrm{KHOBC}}$and $v^{\mathrm{SHOBC}%
}$, respectively. Also, the acronyms CC, CF, CS and SS stand for doubly
clamped, clamped-free, clamped-simply supported and simply supported
boundary conditions, respectively.
\noindent
The following non-dimensional variable $\zeta $, the non-dimensional
characteristic parameters $\lambda $ and $\mu $ as well as the non-dimensional
deflection $\bar{v}$ are also employed for the examples 
\begin{equation}
\zeta =\frac{x}{L},\mathrm{\;\;\;\;\;\;}\lambda =\frac{c}{L},\mathrm{%
\;\;\;\;\;\;}\mu =\frac{l}{L},\mathrm{\;\;\;\;\;\;}\bar{v}=v\frac{K}{%
q_{0}L^{4}}  \label{GrindEQ__7_}
\end{equation}

\subsection{Doubly clamped nano-beam}
\noindent
A doubly clamped nano-beam of length \textit{L} under a uniform load $%
q_{0}$ is first considered. The classical boundary conditions for doubly
clamped nano-beam are given by, 
\begin{equation}
\begin{array}{l}
{v\left( 0\right) =\partial _{x}v\left( 0\right) =0} \\ 
{v\left( L\right) =\partial _{x}v\left( L\right) =0}%
\end{array}
\label{GrindEQ__8_}
\end{equation}%
Employing the set of classical and higher-order boundary conditions to the
flexural deflection of the nano-beam Eq. \eqref{GrindEQ__4_}, the transverse
displacement field $\bar{v}$ and $\bar{v}^{\mathrm{KHOBC}},\bar{v}^{%
\mathrm{SHOBC}}$according to the nonlocal model and the counterpart
theories can be determined as 
\begin{equation}
\begin{array}{l}
{\bar{v}\left( \zeta \right) =\frac{1}{24}\left( -2\zeta ^{3}+\zeta ^{4}-%
\frac{\zeta ^{2}\left( -1+12\lambda ^{2}+24\lambda ^{3}-12\mu ^{2}-24\lambda
\mu ^{2}\right) }{1+2\lambda }+\frac{2\zeta \left( \lambda +6\lambda
^{2}+12\lambda ^{3}-6\mu ^{2}-12\lambda \mu ^{2}\right) }{1+2\lambda }%
\right. } \\ 
\vspace{-4pt}\cr
{\mathrm{\;\;\;\;\;\;\;\;}\left. -2\frac{\left( \mathrm{e}^{\frac{1-\zeta }{%
\mu }}+\mathrm{e}^{\zeta /\mu }-\left( 1+\mathrm{e}^{1/\mu }\right) \right)
\mu \left( -6\lambda ^{2}-12\lambda ^{3}+6\mu ^{2}+\lambda \left( -1+12\mu
^{2}\right) \right) }{\left( -1+\mathrm{e}^{1/\mu }\right) \left( 1+2\lambda
\right) }\right) } \\ 
\vspace{-4pt}\cr
{\bar{v}^{\mathrm{KHOBC}}\left( \zeta \right) =\frac{\mathrm{e}^{-%
\frac{\zeta }{\mu }}}{24\left( -1-2\mu +\mathrm{e}^{1/\mu }\left( -1+2\mu
\right) \right) }\left( 2\left( \mathrm{e}^{1/\mu }+\mathrm{e}^{2\zeta /\mu
}\right) \mu ^{2}-\mathrm{e}^{\zeta /\mu }\left( \zeta ^{2}+2\zeta \mu +2\mu
^{2}\right.\right.}\\
\vspace{-4pt}\cr
{\mathrm{\;\;\;\;\;\;\;\;\;\;\;\;\;\;\;}\left. \left. -\left( 2\zeta ^{3}-\zeta ^{4}\right) \left( 1+2\mu \right) \right)-\mathrm{e}^{\frac{1+\zeta }{%
\mu }}\left( \zeta ^{2}+\left( \zeta ^{4}-2\zeta ^{3}\right) \left( 1-2\mu
\right) -2\zeta \mu +2\mu ^{2}\right) \right) } \\ 
\vspace{-4pt}\cr
{\bar{v}^{\mathrm{SHOBC}}\left( \zeta \right) =\frac{\mathrm{e}^{-\frac{%
\zeta }{\mu }}}{24\left( -1+\mathrm{e}^{1/\mu }\right) }\left( -12\mu
^{3}\left( \mathrm{e}^{1/\mu }+\mathrm{e}^{2\zeta /\mu }\right) -\mathrm{e}%
^{\zeta /\mu }\left( -2\zeta ^{3}+\zeta ^{4}-12\zeta \mu ^{2}-12\mu
^{3}\right. \right. } \\ 
\vspace{-4pt}\cr
{\left. \left. \mathrm{\;\;\;\;\;\;\;\;\;\;\;\;\;\;}+\zeta ^{2}\left( 1+12\mu ^{2}\right) \right)+\mathrm{e}^{\frac{1+\zeta }{%
\mu }}\left( -2\zeta ^{3}+\zeta ^{4}-12\zeta \mu ^{2}+12\mu ^{3}+\zeta
^{2}\left( 1+12\mu ^{2}\right) \right) \right) }%
\end{array}
\label{GrindEQ__9_}
\end{equation}%
The maximum deflection of the doubly clamped nano-beam is attained at the
mid-span as 
\begin{equation}
\begin{array}{l}
{\bar{v}_{\max }=\frac{\left( 1+10\lambda +48\lambda ^{2}+96\lambda
^{3}-48\left( 1+2\lambda \right) \mu ^{2}-32\mu \tanh \left( \frac{1}{4\mu }%
\right) \left( \lambda \left( 1+6\lambda \left( 1+2\lambda \right) \right)
-6\left( 1+2\lambda \right) \mu ^{2}\right) \right) }{384\left( 1+2\lambda
\right) }} \\ [1.5ex]
%\vspace{-12pt}\cr
{\bar{v}_{\max }^{\mathrm{KHOBC}}=\frac{-1+64\mathrm{e}^{\frac{1}{%
2\mu }}\mu ^{2}-2\mu \left( 5+16\mu \right) +\mathrm{e}^{1/\mu }\left(
-1+2\left( 5-16\mu \right) \mu \right) }{384\left( -1-2\mu +\mathrm{e}%
^{1/\mu }\left( -1+2\mu \right) \right) }} \\ [2.0ex]
%\vspace{-12pt}\cr
{\bar{v}_{\max }^{\mathrm{SHOBC}}=\frac{1}{384}-\frac{\mu ^{2}}{8}+\frac{1}{%
2}\mu ^{3}\tanh \left( \frac{1}{4\mu }\right) }%
\end{array}
\label{GrindEQ__10_}
\end{equation}%
It is apparent from Eq. \eqref{GrindEQ__9_} that the non-dimensional
transverse displacements $\bar{v}^{\mathrm{KHOBC}}$ and $\bar{v}^{%
\mathrm{SHOBC}}$ are independent of non-dimensional characteristic parameter $%
\lambda $.

\subsection{Nano-cantilever}
\noindent
A nano-cantilever with length \textit{L} subject to a uniform load$q_{0}$can
be dealt with in a similar way. The classical boundary conditions for
clamped-simply supported nano-beam are written as
\begin{equation}
\begin{array}{l}
{v\left( 0\right) =\partial _{x}v\left( 0\right) =0} \\ 
{M_{\lambda }\left( L\right) =\partial _{x}M_{\lambda }\left( L\right) =0}%
\end{array}
\label{GrindEQ__11_}
\end{equation}%
Utilizing the set of classical and higher-order boundary conditions in the
flexural deflection of the nano-beam Eq. \eqref{GrindEQ__4_}, the transverse
displacement field $\bar{v}$and $\bar{v}^{\mathrm{KHOBC}},\bar{v}^{%
\mathrm{SHOBC}}$ in accordance with the nonlocal model and the counterpart
theories can be determined as, 
\begin{equation}
\begin{array}{l}
{\bar{v}\left( \zeta \right) =\frac{1}{24}\left( -4\zeta ^{3}+\zeta
^{4}+12\zeta \left( \lambda +2\lambda ^{2}-2\mu ^{2}\right) +6\zeta
^{2}\left( 1-2\lambda ^{2}+2\mu ^{2}\right) \right. } \\ 
%\vspace{-8pt}\cr
{\mathrm{\;\;\;\;\;\;\;\;}\left. -\frac{12\mu \left( -\lambda -2\lambda
^{2}+2\mu ^{2}\right) }{-1+\mathrm{e}^{2/\mu }}\left( \mathrm{e}^{\frac{%
2-\zeta }{\mu }}+\mathrm{e}^{\zeta /\mu }-\mathrm{e}^{2/\mu }-1\right)
\right) } \\ 
%\vspace{-8pt}\cr
{\bar{v}^{\mathrm{KHOBC}}\left( \zeta \right) =\frac{1}{24}(-4\zeta
^{3}+\zeta ^{4}+6\left( \zeta ^{2}+2\mu ^{2}\right) \left( 1-2\lambda
^{2}+2\mu ^{2}\right) -\frac{12\mathrm{e}^{\frac{\zeta }{\mu }}\mu
^{2}\left( -1-2\left( -1+\mathrm{e}^{1/\mu }\right) \left( \lambda ^{2}-\mu
^{2}\right) \right) }{-1+\mathrm{e}^{2/\mu }}} \\ 
%\vspace{-8pt}\cr
{\mathrm{\;\;\;\;\;\;\;\;\;\;\;\;\;\;\;}\left. -\frac{12\zeta \mu \left( 1+%
\mathrm{e}^{2/\mu }-2\left( -1+\mathrm{e}^{1/\mu }\right) ^{2}\left( \lambda
^{2}-\mu ^{2}\right) \right) }{-1+\mathrm{e}^{2/\mu }}-\frac{12\mathrm{e}^{%
\frac{1-\zeta }{\mu }}\mu ^{2}\left( 2\left( \lambda ^{2}-\mu ^{2}\right) +%
\mathrm{e}^{1/\mu }\left( 1-2\lambda ^{2}+2\mu ^{2}\right) \right) }{-1+%
\mathrm{e}^{2/\mu }}\right) } \\ 
%\vspace{-8pt}\cr
{\bar{v}^{\mathrm{SHOBC}}\left( \zeta \right) =\frac{\mathrm{e}^{-\frac{%
\zeta }{\mu }}}{24(-1+\mathrm{e}^{2/\mu })}\left( -24\mu ^{3}\left( \mathrm{e%
}^{2/\mu }+\mathrm{e}^{\frac{2\zeta }{\mu }}\right) \right. +\mathrm{e}%
^{\zeta /\mu }\left( 4\zeta ^{3}-\zeta ^{4}+24\zeta \mu ^{2}+24\mu
^{3} +6\zeta ^{2}\left( -1 \right. \right. } \\ 
%\vspace{-8pt}\cr
{\mathrm{\;\;\;\;\;\;\;\;\;\;\;\;\;\;\;} \left. \left. \left. +2\lambda ^{2}-2\mu ^{2}\right) \right)  +\mathrm{e}^{\frac{2+\zeta }{%
\mu }}\left( -4\zeta ^{3}+\zeta ^{4}-24\zeta \mu ^{2}+24\mu ^{3}+\zeta
^{2}\left( 6-12\lambda ^{2}+12\mu ^{2}\right) \right) \right) }%
\end{array}
\label{GrindEQ__12_}
\end{equation}%
which depends on both non-dimensional characteristic parameters$\lambda $and $%
\mu $. Also, the maximum deflection of the nano-beam is attained at the tip
as, 
\begin{equation}
\begin{array}{l}
{\bar{v}_{\max }=\frac{\mu \left( \lambda +2\lambda ^{2}-2\mu ^{2}\right) }{%
1+\mathrm{e}^{\frac{1}{\mu }}}+\frac{1}{8}\left( \left( 1+2\lambda \right)
^{2}-4\lambda \left( 1+2\lambda \right) \mu -4\mu ^{2}+8\mu ^{3}\right) } \\ 
\vspace{-4pt}\cr
{\bar{v}_{\max }^{\mathrm{KHOBC}}=-\frac{1}{8}\left( -2\mu +\coth
\left( \frac{1}{2\mu }\right) \right) \left( \left( -1+4\lambda ^{2}-4\mu
^{2}\right) \tanh \left( \frac{1}{2\mu }\right) +2\mu \right) } \\ 
\vspace{-4pt}\cr
{\bar{v}_{\max }^{\mathrm{SHOBC}}=\frac{1}{8}\left( 1-4\lambda ^{2}-4\mu
^{2}+8\mu ^{3}\tanh \left( \frac{1}{2\mu }\right) \right) }%
\end{array}
\label{GrindEQ__13_}
\end{equation}

\subsection{Clamped-simply supported nano-beam}
\noindent
To examine a clamped-simply supported nano-beam with length \textit{L}
subject to a uniform load $q_{0}$, the classical boundary conditions are
considered as 
\begin{equation}
\begin{array}{l}
{v\left( 0\right) =\partial _{x}v\left( 0\right) =0} \\ 
{v\left( L\right) =M_{\lambda }\left( L\right) =0.}%
\end{array}
\label{GrindEQ__14_}
\end{equation}%
As a result of imposing the set of classical and higher-order boundary
conditions to the flexural deflection of the nano-beam Eq. %
\eqref{GrindEQ__4_}, the transverse displacement field $\bar{v}$ and $\bar{v}%
^{\mathrm{KHOBC}}$, $\bar{v}^{\mathrm{SHOBC}}$ consistent with the
nonlocal model and the counterpart theories can be determined as 

\begin{equation*}
\begin{array}{l}
{\bar{v}\left( \zeta \right) =\frac{-\mathrm{e}^{1/\mu }}{24\left( -1+%
\mathrm{e}^{1/\mu }\right) \left( 1+3\lambda \left( 1+\mu \right) -3\mu
^{2}\left( 1+2\mu \right) +\lambda ^{2}\left( 3+6\mu \right) +\mathrm{e}%
^{1/\mu }\left( 1+3\lambda \left( 1+\lambda \right) -3\lambda \left(
1+2\lambda \right) \mu -3\mu ^{2}+6\mu ^{3}\right) \right) }} \\ 
{\mathrm{\;\;\;\;\;\;\;\;\;}\left( 6\lambda \left( 1+2\lambda \right) \left(
-\left( -2+\zeta \right) \zeta ^{3}+3\lambda +6\left( 1+2\left( -1+\zeta
\right) \zeta \right) \lambda ^{2}\right) \mu \right. } \\ 
{\mathrm{\;\;\;\;\;\;\;\;\;}-6\left( 3+4\zeta ^{3}-2\zeta ^{4}+12\left(
1+\left( -1+\zeta \right) \zeta \right) \lambda +24\left( 1+2\left( -1+\zeta
\right) \zeta \right) \lambda ^{2}\right) \mu ^{3}} \\ 
{\mathrm{\;\;\;\;\;\;\;\;\;}+72\left( 1+2\left( -1+\zeta \right) \zeta
\right) \mu ^{5} -6\mu \cosh \left( \frac{1}{\mu }\right) \left(
-\lambda \left( 1+2\lambda \right) \right.} \\ 
{\mathrm{\;\;\;\;\;\;\;\;\;} \left. \left( 1-2\zeta ^{3}+\zeta ^{4}+3\lambda
+6\left( 1-2\left( -1+\zeta \right) \zeta \right) \lambda ^{2}\right)
\right. } \\ 
{\mathrm{\;\;\;\;\;\;\;\;\;}+\left( 5+2\left( -2+\zeta \right) \zeta
^{3}+12\lambda -12\left( -1+\zeta \right) \zeta \lambda +24\left( 1-2\left(
-1+\zeta \right) \zeta \right) \lambda ^{2}\right) \mu ^{2}} \\ 
{\left. \mathrm{\;\;\;\;\;\;\;\;\;}+12\left( -1+2\left( -1+\zeta \right)
\zeta \right) \mu ^{4}\right)-6\mu \cosh \left( \frac{1-\zeta }{\mu }\right)
\left( \lambda \left( 1+2\lambda \right) \right. } \\ 
{\mathrm{\;\;\;\;\;\;\;\;\;} \left. \left( 1+3\lambda +6\lambda
^{2}\right) -\left( 5+12\lambda \left( 1+2\lambda \right) \right) \mu
^{2}+12\mu ^{4}\right) +\zeta ^{2}\left( -3+\left( 5-2\zeta \right)
\zeta \right) \sinh \left( \frac{1}{\mu }\right) } \\ 
{\mathrm{\;\;\;\;\;\;\;\;\;} -6\sinh \left( \frac{1}{\mu 
}\right) \left( \left( -1+\zeta \right) \zeta \lambda \left( -1-5\lambda
+\left( 1+\lambda \right) \left( \left( -1+\zeta \right) \zeta -12\lambda
^{2}\right) \right) \right. } \\ 
{\mathrm{\;\;\;\;\;\;\;\;\;}+\mu ^{2}\left( -\left( -1+\zeta \right) \zeta
\left( -5+\left( -1+\zeta \right) \zeta \right) +12\left( -1+\zeta \right)
\zeta \lambda +24\left( -1+\zeta \right) \zeta \lambda ^{2} \right. } \\ 
{\left. \mathrm{\;\;\;\;\;\;\;\;\;} \left. +12\lambda
^{3}+24\lambda ^{4}\right)-12\left( \left( -1+\zeta \right) \zeta
+\lambda +4\lambda ^{2}\right) \mu ^{4}+24\mu ^{6}\right) } \\ 
{\mathrm{\;\;\;\;\;\;\;\;\;}+18\mu \left( -\lambda ^{2}+\mu ^{2}\right)
\cosh \left( \frac{\zeta }{\mu }\right) \left( \left( 1+2\lambda \right)
^{2}-4\mu ^{2}-4\mu \left( \lambda +2\lambda ^{2}-2\mu ^{2}\right) \sinh
\left( \frac{1}{\mu }\right) \right) } \\ 
{\mathrm{\;\;\;\;\;\;\;\;\;}\left. -144\left( \lambda ^{2}-\mu ^{2}\right)
\mu ^{2}\left( \lambda +2\lambda ^{2}-2\mu ^{2}\right) \sinh \left( \frac{1}{%
2\mu }\right) ^{2}\sinh \left( \frac{\zeta }{\mu }\right) \right) } \\ [2.0ex]
{\bar{v}^{\mathrm{KHOBC}}\left( \zeta \right) =\frac{\mathrm{e}%
^{1/\mu }}{24\left( -1-3\mu \left( 1+\mu \right) +\mathrm{e}^{2/\mu }\left(
1+3\left( -1+\mu \right) \mu \right) \right) }} \\ 
{\mathrm{\;\;\;\;\;\;\;\;\;\;\;\;\;\;\;\;}\left( -24\left( -1+\zeta \right)
\left( \lambda ^{2}-\mu ^{2}\right) \mu \left( \left( -2+\zeta \right) \zeta
+6\mu ^{2}\right) \right. } \\ 
{\mathrm{\;\;\;\;\;\;\;\;\;\;\;\;\;\;\;\;}-6\zeta \mu \cosh \left( \frac{1}{%
\mu }\right) \left( 1+4\lambda ^{2}+\zeta ^{2}\left( -2+\zeta -4\lambda
^{2}\right) +4\left( -1+\zeta ^{2}-6\lambda ^{2}\right) \mu ^{2}+24\mu
^{4}\right) } \\ 
{\mathrm{\;\;\;\;\;\;\;\;\;\;\;\;\;\;\;\;}+\sinh \left( \frac{1}{\mu }%
\right) \left( \left( -1+\zeta \right) \zeta ^{2}\left( -3+2\zeta -12\lambda
^{2}\right) +6\left( \left( -1+\zeta ^{2}\right) ^{2}+4\left( 1-3\zeta
^{2}\right) \lambda ^{2}\right) \mu ^{2}\right. } \\ 
{\left. \mathrm{\;\;\;\;\;\;\;\;\;\;\;\;\;\;\;\;}+24\left( -1+3\zeta
^{2}-6\lambda ^{2}\right) \mu ^{4}+144\mu ^{6}\right) +6\mu ^{2}\cosh \left( 
\frac{\zeta }{\mu }\right) \left( 24\mu \left( -\lambda ^{2}+\mu ^{2}\right)
\right. } \\ 
{\left. \mathrm{\;\;\;\;\;\;\;\;\;\;\;\;\;\;\;\;}+\left( -1+4\mu ^{2}-24\mu
^{4}+4\lambda ^{2}\left( -1+6\mu ^{2}\right) \right) \sinh \left( \frac{1}{%
\mu }\right) \right) +6\mu ^{2}\left( 8\left(
\lambda ^{2}-\mu ^{2}\right) \left( 1+3\mu ^{2}\right) \right.  } \\ 
{\left. \mathrm{\;\;\;\;\;\;\;\;\;\;\;\;\;\;\;\;} \left. +\left( 1+4\lambda
^{2}-4\left( 1+6\lambda ^{2}\right) \mu ^{2}+24\mu ^{4}\right) \cosh \left( 
\frac{1}{\mu }\right) \right) \sinh \left( \frac{\zeta }{\mu }\right)
\right) }% 
\end{array}%
\end{equation*}
\begin{equation}
\begin{array}{l}
{\bar{v}^{\mathrm{SHOBC}}\left( \zeta \right) =\frac{\mathrm{e}^{-\frac{%
\zeta }{\mu }}}{48\left( -1+\mathrm{e}^{1/\mu }\right) \left( 1-3\mu
^{2}-6\mu ^{3}+\mathrm{e}^{1/\mu }\left( 1-3\mu ^{2}+6\mu ^{3}\right)
\right) }} \\ 
{\mathrm{\;\;\;\;\;\;\;\;\;\;\;\;\;\;\;}\left( -18\mathrm{e}^{\frac{1+2\zeta 
}{\mu }}\mu ^{3}\left( 1-4\lambda ^{2}-4\mu ^{2}+8\mu ^{3}\right) +18\mathrm{%
e}^{\frac{1}{\mu }}\mu ^{3}\left( -1+4\lambda ^{2}+4\mu ^{2}+8\mu
^{3}\right) \right. } \\ 
{\mathrm{\;\;\;\;\;\;\;\;\;\;\;\;\;\;\;}-6\mathrm{e}^{2/\mu }\mu ^{3}\left(
5+12\lambda ^{2}-12\mu ^{2}+24\mu ^{3}\right) +6\mathrm{e}^{\frac{2\zeta }{%
\mu }}\mu ^{3}\left( -5-12\lambda ^{2}+12\mu ^{2}+24\mu ^{3}\right) } \\ 
{\mathrm{\;\;\;\;\;\;\;\;\;\;\;\;\;\;\;}-12\mathrm{e}^{\frac{1+\zeta }{\mu }%
}\mu ^{3}\left( -4\zeta ^{3}+2\zeta ^{4}-24\zeta \mu ^{2}-24\zeta ^{2}\left(
\lambda ^{2}-\mu ^{2}\right) +3\left( -1+4\lambda ^{2}+4\mu ^{2}\right)
\right) } \\ 
{\mathrm{\;\;\;\;\;\;\;\;\;\;\;\;\;\;\;}+\mathrm{e}^{\frac{2+\zeta }{\mu }%
}\left( 2\zeta ^{4}\left( 1-3\mu ^{2}+6\mu ^{3}\right) -\left( \zeta
^{3}+6\zeta \mu ^{2}+6\mu ^{3}\right) \left( 5+12\lambda ^{2}-12\mu
^{2}+24\mu ^{3}\right) \right. } \\ 
{\left. \mathrm{\;\;\;\;\;\;\;\;\;\;\;\;\;\;\;}+3\zeta ^{2}\left( 1+8\mu
^{2}-24\mu ^{4}+48\mu ^{5}+\lambda ^{2}\left( 4+24\mu ^{2}-48\mu ^{3}\right)
\right) \right) } \\ 
{\mathrm{\;\;\;\;\;\;\;\;\;\;\;\;\;\;\;}\mathrm{+e}^{\zeta /\mu }\left(
\left( \zeta ^{3}+6\mu ^{3}+6\zeta \mu ^{2}\right) \left( 5+12\lambda
^{2}-12\mu ^{2}-24\mu ^{3}\right) \right. +2\zeta ^{4}\left( -1+3\mu
^{2}+6\mu ^{3}\right) } \\ 
{\mathrm{\;\;\;\;\;\;\;\;\;\;\;\;\;\;}\left. \left. +3\zeta ^{2}\left(
1+8\mu ^{2}-24\mu ^{4}-48\mu ^{5}+4\lambda ^{2}\left( 1+6\mu ^{2}+12\mu
^{3}\right) \right) \right) \right) }% 
\end{array}
\label{GrindEQ__15_}
\end{equation}%
which also depends on both non-dimensional characteristic parameters $\lambda $
and $\mu $. The deflection of the nano-beam at the mid-span will be employed
for the next figures 
\begin{equation*}
\begin{array}{l}
{\tilde{v}=\frac{1}{384\left( 1+\mathrm{e}^{\frac{1}{2\mu }}\right) }} \\ 
{\mathrm{\;\;\;}\frac{1}{\left( 1+3\lambda \left( 1+\mu \right) -3\left( \mu
^{2}-\lambda ^{2}\right) \left( 1+2\mu \right) +\mathrm{e}^{1/\mu }\left(
1+3\lambda \left( 1+\lambda \right) -3\lambda \left( 1+2\lambda \right) \mu
-3\mu ^{2}+6\mu ^{3}\right) \right) }} \\ 
{\mathrm{\;\;\;}\left( \mathrm{2}+\mathrm{e}^{\frac{1}{2\mu }}\left(
2+3\left( \lambda \left( 5+3\lambda \left( 7+16\lambda \left( 1+\lambda
\right) \right) \right) -3\lambda \left( 1+2\lambda \right) \left(
1+16\lambda \left( 1+\lambda \right) \right) \mu \right. \right. \right. }
\\ 
{\mathrm{\;\;\;}-3\left( 7+16\lambda \left( 1+2\lambda \right) \left(
1+4\lambda ^{2}\right) \right) \mu ^{2}+6\left( 9+8\lambda \left( 3+4\lambda
\right) \right) \mu ^{3}} \\ 
{\mathrm{\;\;\;}\left. \left. +48\left( 1+4\lambda \left( 1+4\lambda \right)
\right) \mu ^{4}-96\mu ^{5}-384\mu ^{6}\right) \right) } \\ 
{\mathrm{\;\;\;}+\mathrm{e}^{1/\mu }\left( 2+3\left( \lambda \left(
5+3\lambda \left( 7+16\lambda \left( 1+\lambda \right) \right) \right)
+3\lambda \left( 1+2\lambda \right) \left( 1+16\lambda \left( 1+\lambda
\right) \right) \mu \right. \right. } \\ 
{\mathrm{\;\;\;}-3\left( 7+16\lambda \left( 1+2\lambda \right) \left(
1+4\lambda ^{2}\right) \right) \mu ^{2}-6\left( 9+8\lambda \left( 3+4\lambda
\right) \right) \mu ^{3}} \\ 
{\mathrm{\;\;\;}\left. \left. +48\left( 1+4\lambda \left( 1+4\lambda \right)
\right) \mu ^{4}+96\mu ^{5}-384\mu ^{6}\right) \right) } \\ 
{\mathrm{\;\;\;}+\mathrm{e}^{\frac{3}{2\mu }}\left( 2+3\left( \lambda \left(
5+3\lambda \left( 7+16\lambda \left( 1+\lambda \right) \right) \right)
-\lambda \left( 1+2\lambda \right) \left( 13+48\lambda \left( 1+3\lambda
\right) \right) \mu \right. \right. } \\ 
{\mathrm{\;\;\;}+3\left( -7+16\lambda \left( -1+2\lambda \right) \left(
1+2\lambda \right) ^{2}\right) \mu ^{2}-2\left( 37+24\lambda \left(
5+12\lambda \right) \right) \mu ^{3}} \\ 
{\mathrm{\;\;\;}\left. \left. -48\left( -1+4\lambda \left( 1+4\lambda
\right) \right) \mu ^{4}-288\mu ^{5}+384\mu ^{6}\right) \right) +3\left( 48\lambda ^{3}\left( 1+\lambda \right) \left(1+4\lambda \right) \right. } \\ 
{\mathrm{\;\;\;} \left. +48\lambda ^{4}\left( 1+6\mu +8\mu ^{2}\right) +\lambda
\left( 5-\mu \left( -13+48\mu \left( 1+\mu \right) \left( 1+4\mu \right)
\right) +\mu ^{2}\left( -21+2\mu \left( -37 \right) \right. \right. \right. } \\ 
{\mathrm{\;\;\;}\left. \left. \left. \left. +24\mu
\left( 1+2\mu \right) \left( 1+4\mu \right) \right) \right) +\lambda
^{2}\left( 21-2\mu \left( -37+48\mu \left( 1+2\mu \right) \left( 1+4\mu
\right) \right) \right) \right) \right) }% 
\end{array}%
\end{equation*}%
\begin{equation}
\begin{array}{l}
{\tilde{v}^{\mathrm{KHOBC}}=\frac{\mathrm{e}^{1/\mu }}{192\left(
-1-3\mu \left( 1+\mu \right) +\mathrm{e}^{2/\mu }\left( 1+3\left( -1+\mu
\right) \mu \right) \right) }} \\ 
{\mathrm{\;\;\;\;\;\;\;\;\;}\left( 72\left( \lambda ^{2}-\mu ^{2}\right) \mu
\left( -1+8\mu ^{2}\right) -1152\mu ^{3}\left( \lambda ^{2}-\mu ^{2}\right)
\cosh \left( \frac{1}{2\mu }\right) \right. } \\ 
{\mathrm{\;\;\;\;\;\;\;\;\;}-3\mu \left( 5+24\lambda ^{2}-24\left(
1+8\lambda ^{2}\right) \mu ^{2}+192\mu ^{4}\right) \cosh \left( \frac{1}{\mu 
}\right) } \\ 
{\mathrm{\;\;\;\;\;\;\;\;\;}-48\mu ^{2}\left( 1+4\mu ^{2}+48\mu
^{4}-4\lambda ^{2}\left( 1+12\mu ^{2}\right) \right) \sinh \left( \frac{1}{%
2\mu }\right) } \\ 
{\mathrm{\;\;\;\;\;\;\;\;\;}\left. +\left( 2+27\mu ^{2}+12\left( -4\mu
^{4}+96\mu ^{6}+\lambda ^{2}\left( 1+4\mu ^{2}-96\mu ^{4}\right) \right)
\right) \sinh \left( \frac{1}{\mu }\right) \right) } \\ [2.0ex]
{\tilde{v}^{\mathrm{SHOBC}}=\frac{1}{384}\left( 5+48\lambda ^{2}-48\mu
^{2}+192\mu ^{3}\right. -\frac{384\mu ^{3}}{1+\mathrm{e}^{\frac{1}{2\mu }}}-%
\frac{3+36\lambda ^{2}}{1-3\mu ^{2}+6\mu ^{3}}} \\ 
{\mathrm{\;\;\;\;\;\;\;\;\;}\left. -\frac{36\left( 1+12\lambda ^{2}\right)
\mu ^{3}}{1-6\mu ^{2}+9\mu ^{4}-36\mu ^{6}+\mathrm{e}^{1/\mu }\left( 1-3\mu
^{2}+6\mu ^{3}\right) ^{2}}\right) }%
\end{array}
\label{GrindEQ__16_}
\end{equation}

\subsection{Simply supported nano-beam}
\noindent
Finally, a simply supported nano-beam with length $L$ subject to a uniform
load $q_{0}$ is examined where the classical boundary conditions are
considered as 
\begin{equation}
\begin{array}{l}
{v\left( 0\right) =M_{\lambda }\left( 0\right) =0} \\ 
{v\left( L\right) =M_{\lambda }\left( L\right) =0}%
\end{array}
\label{GrindEQ__17_}
\end{equation}%
In consequence of imposing the set of classical and higher-order boundary
conditions to the flexural deflection of the nano-beam Eq. %
\eqref{GrindEQ__4_}, the transverse displacement field $\bar{v}$ and $\bar{v}%
^{\mathrm{KHOBC}}$, $\bar{v}^{\mathrm{SHOBC}}$ consistent with the
nonlocal model and the counterpart theories can be determined as 
\begin{equation}
\begin{array}{l}
{\bar{v}\left( \zeta \right) =\frac{\mathrm{e}^{-\frac{\zeta }{\mu }}}{%
24\left( -1+\mathrm{e}^{1/\mu }\right) }\left( -12\left( \mathrm{e}^{\frac{1%
}{\mu }}+\mathrm{e}^{\frac{2\zeta }{\mu }}\right) \mu (-\lambda ^{2}+\mu
^{2})\right. } \\ 
{\mathrm{\;\;\;\;\;\;\;\;}-\mathrm{e}^{\zeta /\mu }\left( -2\zeta ^{3}+\zeta
^{4}+\zeta \left( 1+12\lambda ^{2}-12\mu ^{2}\right) -12\left( \zeta
^{2}-\mu \right) \left( \lambda ^{2}-\mu ^{2}\right) \right) } \\ 
{\mathrm{\;\;\;\;\;\;\;\;}\left. +\mathrm{e}^{\frac{1+\zeta }{\mu }}\left(
-2\zeta ^{3}+\zeta ^{4}+\zeta \left( 1+12\lambda ^{2}-12\mu ^{2}\right)
-12\left( \zeta ^{2}+\mu \right) \left( \lambda ^{2}-\mu ^{2}\right) \right)
\right) } \\ 
{\bar{v}^{\mathrm{KHOBC}}\left( \zeta \right) =\frac{1}{24}\left(
-2\zeta ^{3}+\zeta ^{4}+\zeta \left( 1+12\lambda ^{2}-12\mu ^{2}\right)
-12\zeta ^{2}\left( \lambda ^{2}-\mu ^{2}\right) \right. } \\ 
{\mathrm{\;\;\;\;\;\;\;\;\;\;\;\;\;\;\;}\left. +\frac{24\mathrm{e}^{-^{\zeta
/\mu }}\left( -1+\mathrm{e}^{\zeta /\mu }\right) \left( -\mathrm{e}^{1/\mu }+%
\mathrm{e}^{\zeta /\mu }\right) \mu ^{2}\left( \lambda ^{2}-\mu ^{2}\right) 
}{1+\mathrm{e}^{1/\mu }}\right) } \\ 
{\bar{v}^{\mathrm{SHOBC}}\left( \zeta \right) =\frac{1}{24\left( -1+\mathrm{%
e}^{1/\mu }\right) }\mathrm{e}^{-\frac{\zeta }{\mu }}\left( -12\mathrm{e}%
^{1/\mu }\mu ^{3}-12\mathrm{e}^{2\zeta /\mu }\mu ^{3}\right. } \\ 
{\mathrm{\;\;\;\;\;\;\;\;\;\;\;\;\;\;\;}-\mathrm{e}^{\zeta /\mu }\left(
-2\zeta ^{3}+\zeta ^{4}-12\mu ^{3}+\zeta \left( 1+12\lambda ^{2}-12\mu
^{2}\right) -12\zeta ^{2}\left( \lambda ^{2}-\mu ^{2}\right) \right) } \\ 
{\mathrm{\;\;\;\;\;\;\;\;\;\;\;\;\;\;\;}\left. +\mathrm{e}^{\frac{1+\zeta }{%
\mu }}\left( -2\zeta ^{3}+\zeta ^{4}+12\mu ^{3}+\zeta \left( 1+12\lambda
^{2}-12\mu ^{2}\right) -12\zeta ^{2}\left( \lambda ^{2}-\mu ^{2}\right)
\right) \right) }%
\end{array}
\label{GrindEQ__18_}
\end{equation}%
which again depends on both non-dimensional characteristic parameters $\lambda 
$ and $\mu $. The maximum deflection of the nano-beam is also achieved at
the mid-span as 
\begin{equation}
\begin{array}{l}
{\bar{v}_{\max }=\frac{1}{384}\left( 5+48\lambda ^{2}-48\mu ^{2}+192\mu
\left( -\lambda ^{2}+\mu ^{2}\right) \tanh \left( \frac{1}{4\mu }\right)
\right) } \\ 
\vspace{-8pt}\cr
{\bar{v}_{\max }^{\mathrm{KHOBC}}=\frac{1}{384}\left( 5+48\lambda
^{2}-48\mu ^{2}+384\left( \lambda ^{2}-\mu ^{2}\right) \mu ^{2}\left( -1+%
\mathrm{sech}\left( \frac{1}{2\mu }\right) \right) \right) } \\ 
\vspace{-8pt}\cr
{\bar{v}_{\max }^{\mathrm{SHOBC}}=\frac{1}{384}\left( 5+48\lambda
^{2}-48\mu ^{2}+192\mu ^{3}\tanh \left( \frac{1}{4\mu }\right) \right) }%
\end{array}
\label{GrindEQ__19_}
\end{equation}

\section{Results and discussion}
\noindent
The proposed nonlocal model and counterpart nonlocal models are furthermore
adapted in order to get the numerical assessment of flexural deflection for
Bernoulli-Euler nano-beams. In order to examine the effects of the
characteristic parameters $\lambda $ and $\mu $ on flexural response of
nano-beams, numerical illustrations regarding a Bernoulli-Euler nano-beam
are also presented and discussed for the aforementioned cases.
\noindent
Figs. 1, 2, 3 and 4 depict the plots of the non-dimensional maximum deflection 
$\bar{v}_{\max }$associated with the innovative nonlocal model and the
counterpart nonlocal theory of elasticity $\bar{v}_{\max }^{\mathrm{CBS}-%
\mathrm{I}}$, $\bar{v}_{\max }^{\mathrm{SHOBC}}$ versus the non-dimensional
characteristic parameter $\lambda $, for different values of non-dimensional
characteristic parameter $\mu $, while CC, CF, CS and SS boundary conditions
are imposed, respectively. While the non-dimensional characteristic parameter $%
\lambda $ is ranging in the interval $\left] 0,0.1\right[ $, the
non-dimensional characteristic parameter $\mu $ is assumed to range in the set 
$\left\{ 0^{+},0.04,0.07,0.1\right\} $. It is noticeably deduced from Figs.
1-4 that the innovative nonlocal model exhibits a softening behavior in
terms of the non-dimensional characteristic parameter $\lambda $ and a
hardening behavior in terms of the non-dimensional characteristic parameter $%
\mu $ for different set of boundary conditions. The non-dimensional maximum
deflection consistent with the counterpart nonlocal theory of elasticity $%
\bar{v}_{\max }^{\mathrm{KHOBC}}$, $\bar{v}_{\max }^{\mathrm{SHOBC}%
}$ also exhibits a softening behavior reducing with respect to the
non-dimensional characteristic parameter $\mu $ for different set of boundary
conditions. However, non-dimensional maximum deflections $\bar{v}_{\max }^{%
\mathrm{KHOBC}}$, $\bar{v}_{\max }^{\mathrm{SHOBC}}$ are
independent of the non-dimensional characteristic parameter $\lambda $ for
doubly clamped nano-beams. While non-dimensional maximum deflections $\bar{v}%
_{\max }^{\mathrm{KHOBC}}$, $\bar{v}_{\max }^{\mathrm{SHOBC}}$ of
nano-cantilevers are reducing with respect to the non-dimensional characteristic
parameter $\lambda $ for nano-cantilevers, non-dimensional maximum deflections
are increasing with respect to the non-dimensional characteristic parameter $%
\lambda $ for clamped-simply supported and simply supported nano-beams.
\noindent
In case of counterpart nonlocal theories of elasticity, the size-effect on
the deflection $\bar{v}_{\max }^{\mathrm{%
KHOBC}}$ associated with KHOBC is more noticeable in comparison 
to the deflection $\bar{v}_{\max }^{\mathrm{SHOBC}}$ associated with SHOBC. While
non-dimensional maximum deflection in accordance to the counterpart nonlocal
theory of elasticity $\bar{v}_{\max }^{\mathrm{KHOBC}}$, $\bar{v}%
_{\max }^{\mathrm{SHOBC}}$ exhibits a hardening or softening or independent
behavior with respect to the non-dimensional characteristic parameter $\lambda 
$, the hardening effect is more noticeable for simply supported nano-beams.
The deflections of nano-beams according to the innovative nonlocal theory and
the counterpart nonlocal theory employing SHOBC coincide when the
non-dimensional characteristic parameter approaches zero $\lambda \rightarrow
0^{+}$, for all values of non-dimensional characteristic parameter $\mu $.
While all size-dependent theories coincide with the results of the local
Bernoulli-Euler beam theory for vanishing non-dimensional characteristic
parameters $\lambda $ and $\mu \rightarrow 0^{+}$, discrepancy between the
flexural results of size-dependent models is enhanced by increasing values
of the non-dimensional characteristic parameters. Finally, all of the
size-dependent models reveal a hardening behavior in terms of the number of
kinematic boundary constraints. The stiffest structural response with
respect to the considered boundary conditions is revealed by doubly clamped
Bernoulli-Euler nano-beams.
\noindent
Also, the numerical results of non-dimensional maximum deflection $\bar{v}%
_{\max }\left( \lambda ,\mu \right) $ evaluated by the innovative nonlocal
model and the counterpart nonlocal theories of elasticity together with the
ratios $\Delta ^{\mathrm{KHOBC}}$ and $\Delta ^{\mathrm{SHOBC}}$
between the gaps of $\bar{v}_{\max }$ of innovative nonlocal model with
respect to the counterpart nonlocal theories of elasticity $\bar{v}_{\max }^{%
\mathrm{KHOBC}}$, $\bar{v}_{\max }^{\mathrm{SHOBC}}$ are reported
in Tables 1, 2, 3 and 4 for CC, CF, CS and SS boundary conditions.

\section{Concluding remarks}

The outcomes of the present paper may be summarized as follows.
\begin{itemize}
\item
Scale phenomena in functionally graded (FG) nano-beams
have been modeled by the nonlocal strain gradient elasticity theory 
proposed in \cite{Lim}, equipped with the innovative natural 
constitutive boundary conditions (CBC) contributed 
in \cite{Barretta18d}. 
\item
The elastostatic problem of a nonlocal strain gradient inflected nano-beam
has been formulated by considering both the CBC and the
higher-order boundary conditions commonly adopted in literature.
Analytical and numerical displacement solutions for 
doubly clamped, clamped-free, clamped-pinned and pinned-pinned
nano-beams subject to a uniformly distributed transversal loading 
have been established and compared.
New benchmarks for numerical analyses have been also detected.
\item
It has been underlined that the nonlocal strain gradient displacement solutions 
exhibit softening and stiffening behaviors
for increasing values of the nonlocal and strain gradient parameters respectively. 
The proposed model is therefore able to capture significantly
size-dependent responses of FG nano-structures.

\end{itemize}

\begin{table}
\begin{tabular}{lllllll}
\hline
\multicolumn{7}{l}{$\bar{v}_{\max }^{{}}$} \\ \hline
$\mu $ & $\lambda $ & $\bar{v}_{\max }^{{}}$ & $\bar{v}_{\max }^{\mathrm{KHOBC}%
}$ & $\Delta ^{\mathrm{KHOBC}}(\%)$ & $\bar{v}_{\max }^{%
\mathrm{SHOBC}}$ & $\Delta ^{\mathrm{SHOBC}}(\%)$ \\ \hline
$0^{+}$ & $0^{+}$ & $0.002604$ & $0.002602$ & $-0.07994$ & $0.002604$ & $0$
\\ \hline
& $0.02$ & $0.003055$ & $0.002602$ & $-14.815$ & $0.002604$ & $-14.7468$ \\ 
\hline
& $0.04$ & $0.003575$ & $0.002602$ & $-27.2222$ & $0.002604$ & $-27.164$ \\ 
\hline
& $0.06$ & $0.00417$ & $0.002602$ & $-37.5941$ & $0.002604$ & $-37.5442$ \\ 
\hline
& $0.08$ & $0.00484$ & $0.002602$ & $-46.2385$ & $0.002604$ & $-46.1955$ \\ 
\hline
& $0.1$ & $0.005589$ & $0.002602$ & $-53.4435$ & $0.002604$ & $-53.4062$ \\ 
\hline
$0.04$ & $0^{+}$ & $0.002436$ & $0.001843$ & $-24.3362$ & $0.002436$ & $0$
\\ \hline
& $0.02$ & $0.002815$ & $0.001843$ & $-34.5119$ & $0.002436$ & $-13.4486$ \\ 
\hline
& $0.04$ & $0.003252$ & $0.001843$ & $-43.3236$ & $0.002436$ & $-25.0944$ \\ 
\hline
& $0.06$ & $0.003752$ & $0.001843$ & $-50.8673$ & $0.002436$ & $-35.0645$ \\ 
\hline
& $0.08$ & $0.004315$ & $0.001843$ & $-57.2823$ & $0.002436$ & $-43.5428$ \\ 
\hline
& $0.1$ & $0.004945$ & $0.001843$ & $-62.7203$ & $0.002436$ & $-50.7298$ \\ 
\hline
$0.07$ & $0^{+}$ & $0.002163$ & $0.001382$ & $-36.0816$ & $0.002163$ & $0$
\\ \hline
& $0.02$ & $0.002488$ & $0.001382$ & $-44.4239$ & $0.002163$ & $-13.0514$ \\ 
\hline
& $0.04$ & $0.002863$ & $0.001382$ & $-51.7099$ & $0.002163$ & $-24.4504$ \\ 
\hline
& $0.06$ & $0.003291$ & $0.001382$ & $-57.9939$ & $0.002163$ & $-34.2817$ \\ 
\hline
& $0.08$ & $0.003774$ & $0.001382$ & $-63.3717$ & $0.002163$ & $-42.6952$ \\ 
\hline
& $0.1$ & $0.004314$ & $0.001382$ & $-67.9551$ & $0.002163$ & $-49.8659$ \\ 
\hline
$0.1$ & $0^{+}$ & $0.001847$ & $0.001028$ & $-44.3617$ & $0.001847$ & $0$ \\ 
\hline
& $0.02$ & $0.00212$ & $0.001028$ & $-51.5203$ & $0.001847$ & $-12.8662$ \\ 
\hline
& $0.04$ & $0.002436$ & $0.001028$ & $-57.7974$ & $0.001847$ & $-24.1483$ \\ 
\hline
& $0.06$ & $0.002796$ & $0.001028$ & $-63.2301$ & $0.001847$ & $-33.9126$ \\ 
\hline
& $0.08$ & $0.003202$ & $0.001028$ & $-67.8933$ & $0.001847$ & $-42.2938$ \\ 
\hline
& $0.1$ & $0.003655$ & $0.001028$ & $-71.8778$ & $0.001847$ & $-49.4553$ \\ 
\hline
\end{tabular}
\caption{Maximum non-dimensional deflection $\bar{v}_{\max }$ 
of CC nano-beams vs. scale parameters $\lambda $, $\mu $}
\end{table}

%\newpage

%%\begin{center}
%%\textbf{TABLE 2 - Maximum dimensionless deflection }$\bar{v}_{\max }$ 
%%\textbf{of CF Bernoulli-Euler nano-beam versus the characteristic
%%dimensionless parameters }$\lambda $, $\mu $\textbf{.}

\begin{table}
\begin{tabular}{lllllll}
\hline
\multicolumn{7}{l}{$\bar{v}_{\max }^{{}}$} \\ \hline
$\mu $ & $\lambda $ & $\bar{v}_{\max }^{{}}$ & $\bar{v}_{\max }^{\mathrm{KHOBC}%
}$ & $\Delta ^{\mathrm{KHOBC}}(\%)$ & $\bar{v}_{\max }^{%
\mathrm{SHOBC}}$ & $\Delta ^{\mathrm{SHOBC}}(\%)$ \\ \hline
$0^{+}$ & $0^{+}$ & $0.125$ & $0.12495$ & $-0.03999$ & $0.125$ & $0$ \\ 
\hline
& $0.02$ & $0.135199$ & $0.12475$ & $-7.72854$ & $0.1248$ & $-7.6916$ \\ 
\hline
& $0.04$ & $0.145798$ & $0.12415$ & $-14.8477$ & $0.1242$ & $-14.8136$ \\ 
\hline
& $0.06$ & $0.156797$ & $0.12315$ & $-21.4585$ & $0.1232$ & $-21.4269$ \\ 
\hline
& $0.08$ & $0.168195$ & $0.121751$ & $-27.6135$ & $0.1218$ & $-27.5842$ \\ 
\hline
& $0.1$ & $0.179994$ & $0.119951$ & $-33.3583$ & $0.12$ & $-33.3311$ \\ 
\hline
$0.04$ & $0^{+}$ & $0.124264$ & $0.106536$ & $-14.2664$ & $0.124264$ & $0$
\\ \hline
& $0.02$ & $0.134048$ & $0.106352$ & $-20.6613$ & $0.124064$ & $-7.44808$ \\ 
\hline
& $0.04$ & $0.1442$ & $0.1058$ & $-26.6297$ & $0.123464$ & $-14.38$ \\ \hline
& $0.06$ & $0.15472$ & $0.10488$ & $-32.213$ & $0.122464$ & $-20.848$ \\ 
\hline
& $0.08$ & $0.165608$ & $0.103592$ & $-37.4475$ & $0.121064$ & $-26.8973$ \\ 
\hline
& $0.1$ & $0.176864$ & $0.101936$ & $-42.3648$ & $0.119264$ & $-32.5674$ \\ 
\hline
$0.07$ & $0^{+}$ & $0.122893$ & $0.094557$ & $-23.0575$ & $0.122893$ & $0$
\\ \hline
& $0.02$ & $0.132365$ & $0.094385$ & $-28.6934$ & $0.122693$ & $-7.30707$ \\ 
\hline
& $0.04$ & $0.142181$ & $0.093869$ & $-33.9792$ & $0.122093$ & $-14.1285$ \\ 
\hline
& $0.06$ & $0.152341$ & $0.093009$ & $-38.9468$ & $0.121093$ & $-20.5119$ \\ 
\hline
& $0.08$ & $0.162845$ & $0.091805$ & $-43.6243$ & $0.119693$ & $-26.4988$ \\ 
\hline
& $0.1$ & $0.173693$ & $0.090257$ & $-48.0365$ & $0.117893$ & $-32.1256$ \\ 
\hline
$0.1$ & $0^{+}$ & $0.121$ & $0.084$ & $-30.5784$ & $0.121$ & $0$ \\ \hline
& $0.02$ & $0.13016$ & $0.08384$ & $-35.5869$ & $0.1208$ & $-7.19122$ \\ 
\hline
& $0.04$ & $0.13964$ & $0.08336$ & $-40.3036$ & $0.1202$ & $-13.9216$ \\ 
\hline
& $0.06$ & $0.14944$ & $0.08256$ & $-44.7538$ & $0.1192$ & $-20.2357$ \\ 
\hline
& $0.08$ & $0.15956$ & $0.08144$ & $-48.9597$ & $0.1178$ & $-26.1722$ \\ 
\hline
& $0.1$ & $0.17$ & $0.08$ & $-52.9413$ & $0.116$ & $-31.7649$ \\ \hline
\end{tabular}
\caption{Maximum non-dimensional deflection $\bar{v}_{\max }$ 
of CF nano-beams vs. scale parameters $\lambda $, $\mu $}
\end{table}

%\newpage

%%\textbf{TABLE 3 - Maximum dimensionless deflection }$\bar{v}_{\max }$ 
%%\textbf{of CS Bernoulli-Euler nano-beam versus the characteristic
%%dimensionless parameters }$\lambda $, $\mu $\textbf{.}

\begin{table}
\begin{tabular}{lllllll}
\hline
\multicolumn{7}{l}{$\bar{v}_{\max }^{{}}$} \\ \hline
$\mu $ & $\lambda $ & $\bar{v}_{\max }^{{}}$ & $\bar{v}_{\max }^{\mathrm{KHOBC}%
}$ & $\Delta ^{\mathrm{KHOBC}}(\%)$ & $\bar{v}_{\max }^{%
\mathrm{SHOBC}}$ & $\Delta ^{\mathrm{SHOBC}}(\%)$ \\ \hline
$0^{+}$ & $0^{+}$ & $0.005208$ & $0.005206$ & $-0.04497$ & $0.005208$ & $0$
\\ \hline
& $0.02$ & $0.005709$ & $0.005218$ & $-8.58718$ & $0.005221$ & $-8.54608$ \\ 
\hline
& $0.04$ & $0.006275$ & $0.005256$ & $-16.2364$ & $0.005258$ & $-16.1988$ \\ 
\hline
& $0.06$ & $0.00691$ & $0.005318$ & $-23.027$ & $0.005321$ & $-22.9926$ \\ 
\hline
& $0.08$ & $0.007616$ & $0.005406$ & $-29.0154$ & $0.005408$ & $-28.9839$ \\ 
\hline
& $0.1$ & $0.008396$ & $0.005518$ & $-34.2707$ & $0.005521$ & $-34.2417$ \\ 
\hline
$0.04$ & $0^{+}$ & $0.005006$ & $0.004261$ & $-14.8826$ & $0.005006$ & $0$
\\ \hline
& $0.02$ & $0.005437$ & $0.004272$ & $-21.4379$ & $0.005018$ & $-7.70949$ \\ 
\hline
& $0.04$ & $0.005926$ & $0.004304$ & $-27.3642$ & $0.005055$ & $-14.6933$ \\ 
\hline
& $0.06$ & $0.006473$ & $0.004359$ & $-32.6698$ & $0.005117$ & $-20.9575$ \\ 
\hline
& $0.08$ & $0.007082$ & $0.004435$ & $-37.3832$ & $0.005203$ & $-26.5328$ \\ 
\hline
& $0.1$ & $0.007754$ & $0.004532$ & $-41.5464$ & $0.005314$ & $-31.4657$ \\ 
\hline
$0.07$ & $0^{+}$ & $0.004667$ & $0.003575$ & $-23.3898$ & $0.004667$ & $0$
\\ \hline
& $0.02$ & $0.00505$ & $0.003585$ & $-29.0114$ & $0.004679$ & $-7.34631$ \\ 
\hline
& $0.04$ & $0.005483$ & $0.003614$ & $-34.0934$ & $0.004715$ & $-14.0026$ \\ 
\hline
& $0.06$ & $0.005967$ & $0.003661$ & $-38.6431$ & $0.004775$ & $-19.9744$ \\ 
\hline
& $0.08$ & $0.006504$ & $0.003728$ & $-42.6848$ & $0.004859$ & $-25.2907$ \\ 
\hline
& $0.1$ & $0.007096$ & $0.003814$ & $-46.2543$ & $0.004968$ & $-29.9952$ \\ 
\hline
$0.1$ & $0^{+}$ & $0.00426$ & $0.002958$ & $-30.5464$ & $0.00426$ & $0$ \\ 
\hline
& $0.02$ & $0.004597$ & $0.002967$ & $-35.4653$ & $0.004271$ & $-7.08725$ \\ 
\hline
& $0.04$ & $0.004977$ & $0.002991$ & $-39.8974$ & $0.004306$ & $-13.4821$ \\ 
\hline
& $0.06$ & $0.005401$ & $0.003032$ & $-43.8535$ & $0.004364$ & $-19.1979$ \\ 
\hline
& $0.08$ & $0.005869$ & $0.00309$ & $-47.3584$ & $0.004445$ & $-24.2686$ \\ 
\hline
& $0.1$ & $0.006384$ & $0.003163$ & $-50.4458$ & $0.004549$ & $-28.7412$ \\ 
\hline
\end{tabular}
\caption{Maximum non-dimensional deflection $\bar{v}_{\max }$ 
of CS nano-beams vs. scale parameters $\lambda $, $\mu $}
\end{table}

%\newpage

%%\textbf{TABLE 4 - Maximum dimensionless deflection }$\bar{v}_{\max }$ 
%%\textbf{of SS Bernoulli-Euler nano-beam versus the characteristic
%%dimensionless parameters }$\lambda $, $\mu $\textbf{.}

\begin{table}
\begin{tabular}{lllllll}
\hline
\multicolumn{7}{l}{$\bar{v}_{\max }^{{}}$} \\ \hline
$\mu $ & $\lambda $ & $\bar{v}_{\max }^{{}}$ & $\bar{v}_{\max }^{\mathrm{KHOBC}%
}$ & $\Delta ^{\mathrm{KHOBC}}(\%)$ & $\bar{v}_{\max }^{%
\mathrm{SHOBC}}$ & $\Delta ^{\mathrm{SHOBC}}(\%)$ \\ \hline
$0^{+}$ & $0^{+}$ & $0.013021$ & $0.013021$ & $0$ & $0.013021$ & $0$ \\ 
\hline
& $0.02$ & $0.013071$ & $0.013071$ & $0.000153$ & $0.013071$ & $0.000153$ \\ 
\hline
& $0.04$ & $0.013221$ & $0.013221$ & $0.000605$ & $0.013221$ & $0.000605$ \\ 
\hline
& $0.06$ & $0.013471$ & $0.013471$ & $0.001336$ & $0.013471$ & $0.001336$ \\ 
\hline
& $0.08$ & $0.013821$ & $0.013821$ & $0.002315$ & $0.013821$ & $0.002315$ \\ 
\hline
& $0.1$ & $0.01427$ & $0.014271$ & $0.003503$ & $0.014271$ & $0.003504$ \\ 
\hline
$0.04$ & $0^{+}$ & $0.012853$ & $0.012823$ & $-0.22905$ & $0.012853$ & $0$
\\ \hline
& $0.02$ & $0.012895$ & $0.012873$ & $-0.17123$ & $0.012903$ & $0.06204$ \\ 
\hline
& $0.04$ & $0.013021$ & $0.013021$ & $0$ & $0.013053$ & $0.245758$ \\ \hline
& $0.06$ & $0.013231$ & $0.013268$ & $0.278136$ & $0.013303$ & $0.544179$ \\ 
\hline
& $0.08$ & $0.013525$ & $0.013613$ & $0.653016$ & $0.013653$ & $0.9464$ \\ 
\hline
& $0.1$ & $0.013903$ & $0.014057$ & $1.11171$ & $0.014103$ & $1.43854$ \\ 
\hline
$0.07$ & $0^{+}$ & $0.01258$ & $0.012432$ & $-1.17061$ & $0.01258$ & $0$ \\ 
\hline
& $0.02$ & $0.012616$ & $0.01248$ & $-1.07198$ & $0.01263$ & $0.110799$ \\ 
\hline
& $0.04$ & $0.012724$ & $0.012625$ & $-0.77944$ & $0.01278$ & $0.43943$ \\ 
\hline
& $0.06$ & $0.012904$ & $0.012865$ & $-0.30277$ & $0.01303$ & $0.974917$ \\ 
\hline
& $0.08$ & $0.013156$ & $0.013201$ & $0.34265$ & $0.01338$ & $1.69997$ \\ 
\hline
& $0.1$ & $0.01348$ & $0.013633$ & $1.13699$ & $0.01383$ & $2.59232$ \\ 
\hline
$0.1$ & $0^{+}$ & $0.012264$ & $0.01187$ & $-3.21796$ & $0.012264$ & $0$ \\ 
\hline
& $0.02$ & $0.012294$ & $0.011916$ & $-3.08163$ & $0.012314$ & $0.160498$ \\ 
\hline
& $0.04$ & $0.012385$ & $0.012054$ & $-2.67666$ & $0.012464$ & $0.637285$ \\ 
\hline
& $0.06$ & $0.012537$ & $0.012284$ & $-2.01474$ & $0.012714$ & $1.41658$ \\ 
\hline
& $0.08$ & $0.012748$ & $0.012606$ & $-1.11446$ & $0.013064$ & $2.47651$ \\ 
\hline
& $0.1$ & $0.013021$ & $0.013021$ & $0$ & $0.013514$ & $3.7886$ \\ \hline
\end{tabular}
\caption{Maximum non-dimensional deflection $\bar{v}_{\max }$ 
of SS nano-beams vs. scale parameters $\lambda $, $\mu $}
\end{table}

%%\end{center}

\medskip
\medskip

\textbf{Acknowledgment} - Financial supports from the Italian Ministry of
Education, University and Research (MIUR) in the framework of the Project
PRIN 2015 \textquotedblleft COAN 5.50.16.01\textquotedblright\ - code
2015JW9NJT - and from the research program ReLUIS 2018 are gratefully
acknowledged.

\break

\bigskip

\newpage

\bigskip

\textbf{List of Figures}\bigskip

\begin{itemize}
\item Fig. 1: Maximum non-dimensional deflection $\bar{v}_{\max }$ 
of clamped-clamped (CC) nano-beams vs. scale parameters $\lambda $, $\mu $,
associated with the natural CBC Eq. \eqref{g8} and with higher-order
boundary conditions KHOBC Eq. \eqref{GrindEQ__5_} and SHOBC Eq. \eqref{GrindEQ__6_}.

\item Fig. 2: Maximum non-dimensional deflection $\bar{v}_{\max }$ 
of clamped-free (CF) nano-beams vs. scale parameters $\lambda $, $\mu $,
associated with the natural CBC Eq. \eqref{g8} and with higher-order
boundary conditions KHOBC Eq. \eqref{GrindEQ__5_} and SHOBC Eq. \eqref{GrindEQ__6_}.

\item Fig. 3: Maximum non-dimensional deflection $\bar{v}_{\max }$ 
of clamped-supported (CS) nano-beams vs. scale parameters $\lambda $, $\mu $,
associated with the natural CBC Eq. \eqref{g8} and with higher-order
boundary conditions KHOBC Eq. \eqref{GrindEQ__5_} and SHOBC Eq. \eqref{GrindEQ__6_}.

\item Fig. 4: Maximum non-dimensional deflection $\bar{v}_{\max }$ 
of supported-supported (SS) nano-beams vs. scale parameters $\lambda $, $\mu $,
associated with the natural CBC Eq. \eqref{g8} and with higher-order
boundary conditions KHOBC Eq. \eqref{GrindEQ__5_} and SHOBC Eq. \eqref{GrindEQ__6_}.
\end{itemize}

\begin{figure}[h]
\centering	
\includegraphics{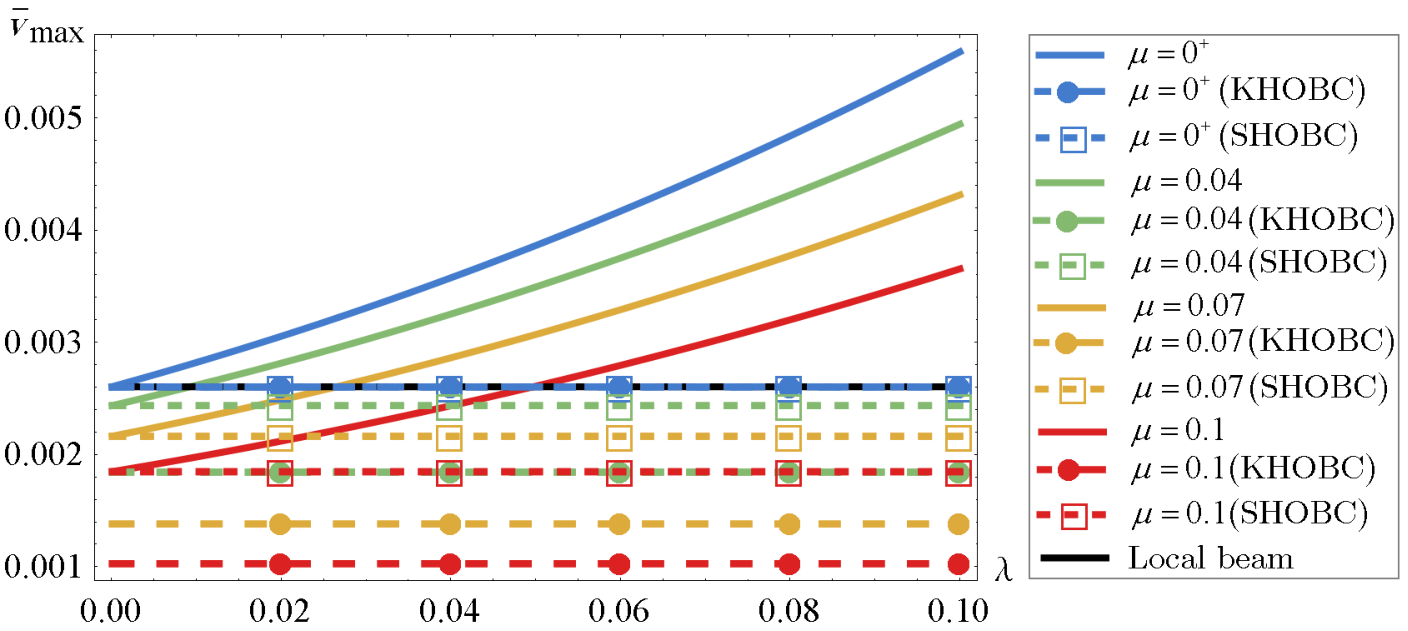}
\caption{}
\label{fig:1}
\end{figure}

\begin{figure}[h]
\centering	
\includegraphics{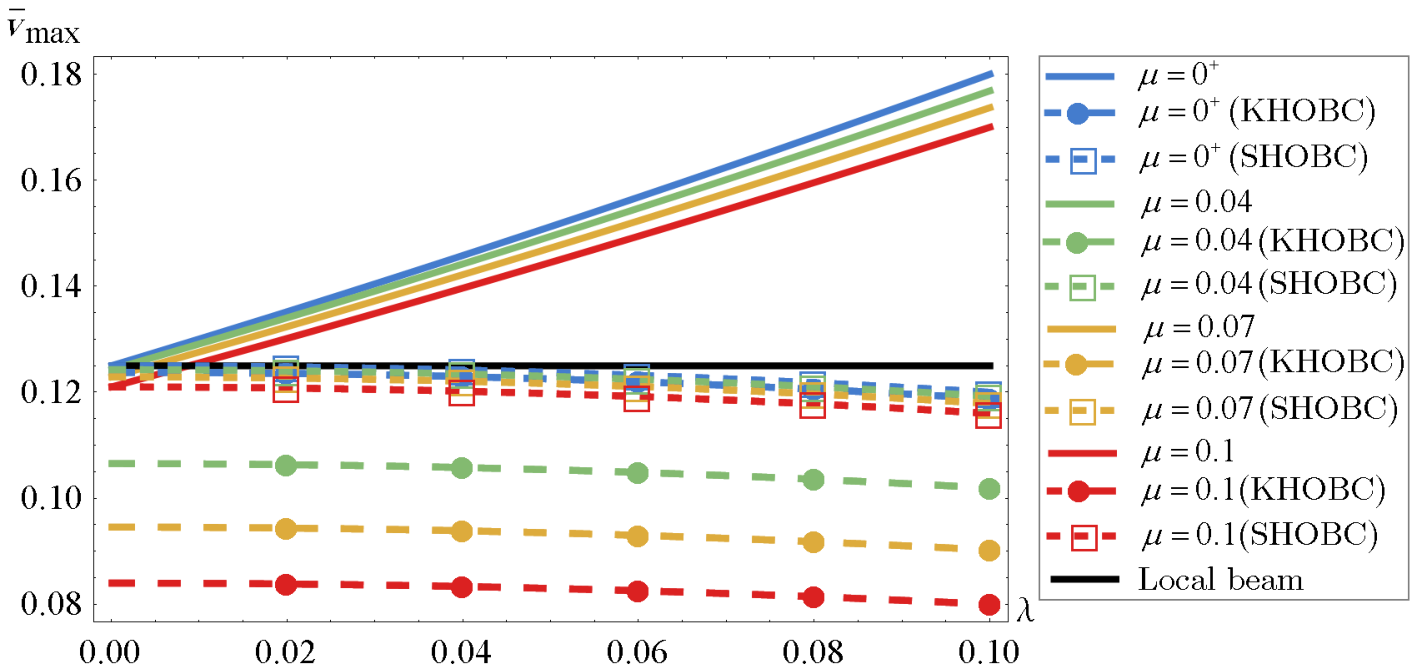}
\caption{}
\label{fig:2}
\end{figure}

\begin{figure}[h]
\centering	
\includegraphics{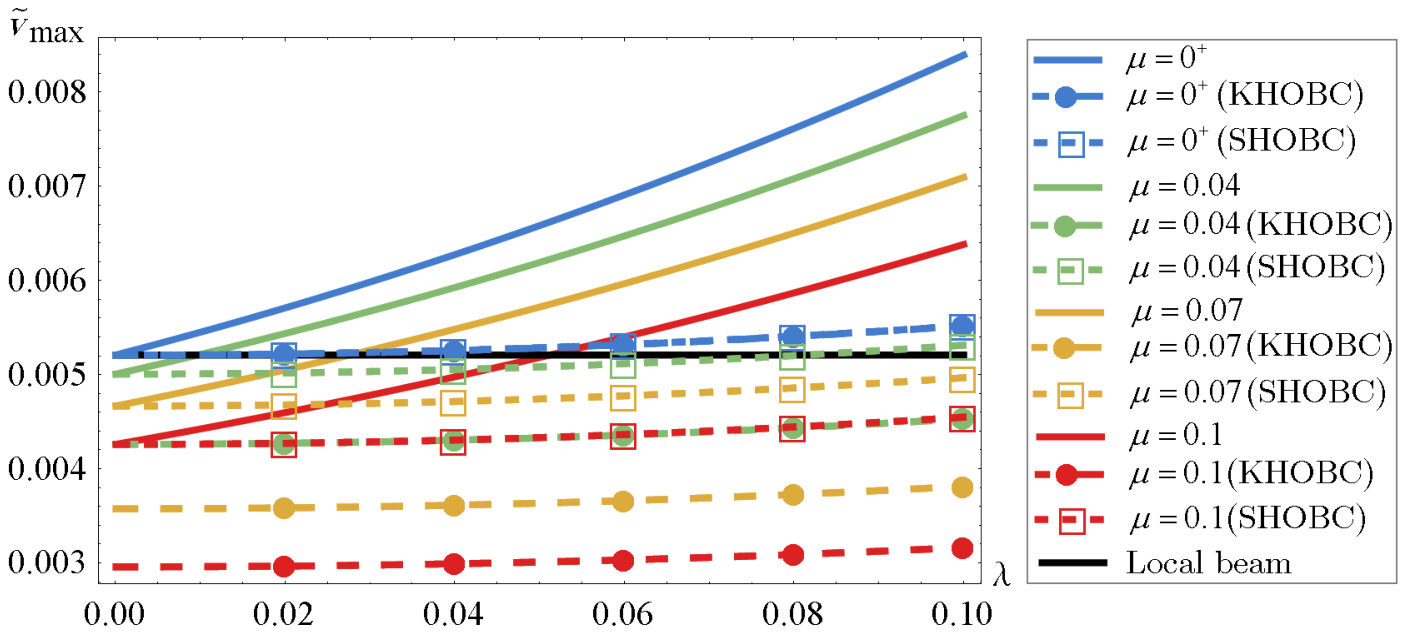}
\caption{}
\label{fig:3}
\end{figure}

\begin{figure}[h]
\centering	
\includegraphics{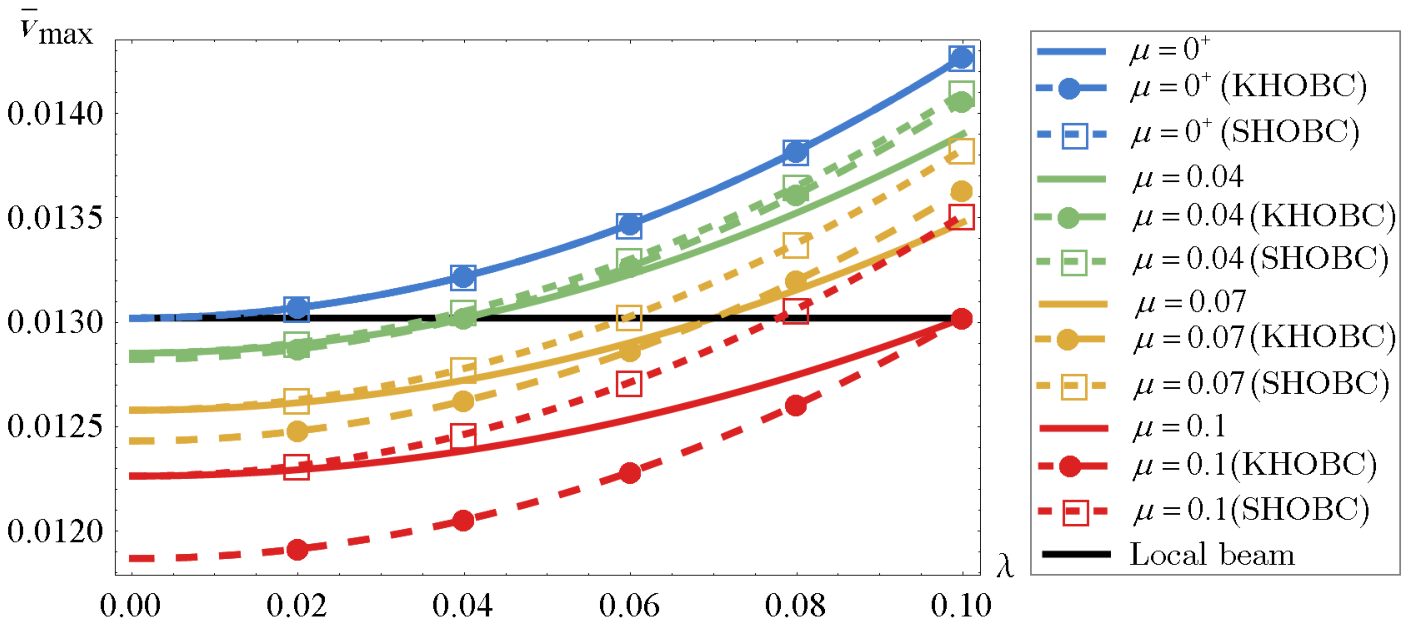}
\caption{}
\label{fig:4}
\end{figure}

\end{document}